# Three-dimensional manipulation of electron beam phase space for seeding soft x-ray free-electron lasers

Chao Feng, Tong Zhang, Haixiao Deng and Zhentang Zhao[a]

*Shanghai Institute of Applied Physics, Chinese Academy of Sciences, Shanghai 201800, China*

In this letter, a simple technique is proposed to induce strong density modulation into the electron beam with small energy modulation. By using the combination of a transversely dispersed electron beam and a wave-front tilted seed laser, three-dimensional manipulation of the electron beam phase space can be utilized to significantly enhance the micro-bunching of seeded free-electron laser schemes, which will improve the performance and extend the short-wavelength range of a single-stage seeded free-electron laser. Theoretical analysis and numerical simulations demonstrate the capability of the proposed technique in a soft x-ray free-electron laser.

Free-electron lasers (FELs) hold great promise as extremely high intensity, coherent photon sources for short-wavelength radiation, which is opening up new frontiers of ultra-fast and ultra-small sciences at the atomic scale. Most of the currently existing FEL facilities in the x-ray wavelength are based on the self-amplified spontaneous emission (SASE) principle[1-4]. While the radiation from a SASE FEL has excellent transverse coherence, it typically has rather limited temporal coherence as the initial radiation starts from the electron beam shot noise. To improve the temporal coherence of a SASE based FEL, several novel techniques, such as self-seeding[5-7], purified-SASE[8], improved-SASE[9], and HB-SASE[10], have been developed recently.

An alternative way for significantly improving the temporal coherence of high-gain FELs is frequency up-conversion schemes, which generally relay on the techniques of precise optical-scale manipulation of the electron beam longitudinal phase space with the help of external coherent laser beams. In the high-gain harmonic generation (HGHG) scheme[11], typically a seed laser pulse is first used to interact with the electrons in a short undulator, called modulator, to generate a sinusoidal energy modulation in the electron beam longitudinal phase space at the scale of seed laser wavelength. The formed energy modulation then develops into an associated density modulation by a dispersive magnetic chicane, called the dispersion section (DS). High harmonic components of the seed laser frequency can be generated in the electron beam density distribution, making this approach powerful for generating short wavelength FEL pulses. Generally, the output properties of HGHG is a direct map of the seed laser's attributes, which ensures high degree of temporal coherence and small pulse energy fluctuations with respect to SASE. The theoretical predictions have been demonstrated in worldwide HGHG experiments[12-15]. However, significant bunching at higher harmonics by strengthening the energy modulation would increase the energy spread of the electron beam, which results in a degradation of the amplification process in the radiator. The requirement of FEL amplification on the beam energy spread prevents the possibility of reaching short wavelength in a single stage HGHG. Higher harmonics can be obtained by cascaded HGHG with fresh-bunch technique[16], which has been demonstrated both at Shanghai SDUV-FEL[17] and Trieste FERMI[18] recently. FERMI is the first seeded FEL facility that works in the x-ray regime, which will open up investigations of many new areas of sciences. In order to improve the frequency multiplication efficiency in a single stage, more complicated phase space manipulation techniques have been developed, e.g., the echo-enabled harmonic generation[19, 20] employs two pairs of modulator and dispersion section to introduce the echo effect into the electron beam phase space for enhancing the frequency multiplication of the current modulation with a relatively small energy modulation. However, all the above mentioned seeded FEL schemes are focused on the longitudinal phase space manipulation. Recently, a novel seeded FEL scheme named phase-merging enhanced harmonic generation (PEHG)[21, 22], has been proposed for significantly improving the frequency up-conversion efficiency of harmonic generation FELs by using a transversely dispersed electron beam, meanwhile in which a transverse gradient undulator is introduced for the energy modulation and the phase-merging effect purpose. Analytical and numerical investigations demonstrate the potential of generating ultra-high harmonic radiation with a relatively small energy modulation in a single stage PEHG.

---

[a] Email: zhaozt@sinap.ac.cn

Here, inspired by these earlier works, we propose a novel method for precisely manipulate the electron beam phase space in three-dimensional, i.e. both longitudinal and transversal, by using the interaction between a wave-front tilted seed laser and a transversely dispersed electron beam in a conventional modulator. It is found that, the effective slice energy spread of the transversely dispersed electron beam can be greatly reduced by simply tilting the wave-front of the seed laser, and thus the phase-merging effect can be easily achieved. Compared with the previous phase space manipulation methods, this new technique is quite easy to implement on all existing seeded FEL facilities.

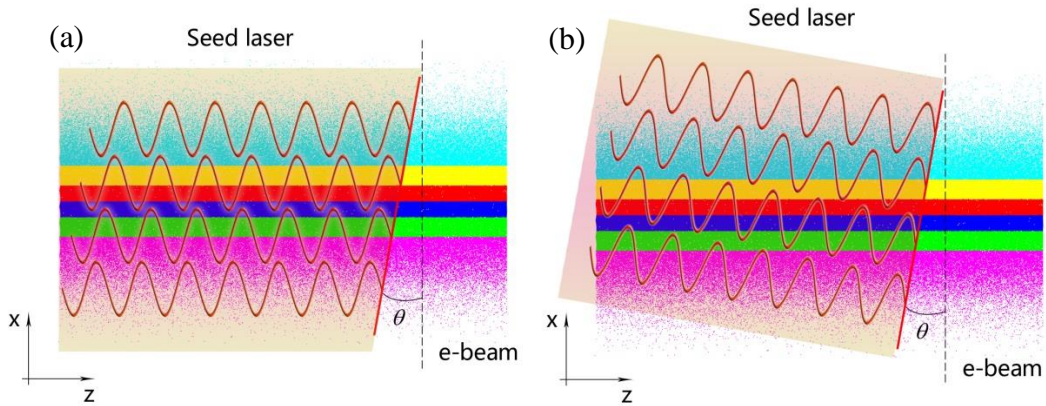

FIG.1. Schematic of two methods for three dimensional manipulations of the electron beam phase space with a transversely dispersed electron beam (x-z) and a wave-front tilted seed laser: (a) directly tilt the wave-front of the seed laser, and (b) oblique incidence the seed laser with a small angle $\theta$.

The schematic of three-dimensional manipulation of the electron beam phase space is shown in Fig. 1, where the electron beam is transversely dispersed before entering the modulator, thus different colors in the electron beam represent different horizontal positions of the electrons with different regions of initial beam energy. The wave-front tilt of the seed laser can be generated by different methods. For example, one could tilt the wave-front by using a grating to reflect the laser beam (Fig. 1(a)) or simply oblique incidence the seed laser beam with a small angle $\theta$ (Fig. 1(b)).

Here we assume an initial Gaussian beam energy distribution with an average energy $\gamma_0 mc^2$ and use the variable $p = (\gamma - \gamma_0)/\sigma_\gamma$ for the dimensionless energy deviation of a particle, where $\sigma_\gamma$ is the initial beam energy spread. Thus the initial longitudinal phase space distribution should be written as $f_0(p) = N_0 \exp(-p^2/2)/\sqrt{2\pi}$, where $N_0$ is the number of electrons per unit length of the beam. Assuming the intrinsic horizontal and vertical rms beam size is $\sigma_x$ and $\sigma_y$, respectively, and use $X = (x-x_0)/\sigma_x$ and $Y = (y-y_0)/\sigma_y$ for the dimensionless horizontal and vertical position of a particle, then the transverse electron beam distribution can be written as $g_0(X,Y) = N_0 \exp(-X^2/2)\exp(-Y^2/2)/2\pi$. After transversely dispersing the electron beam by the horizontal and vertical dispersions $\eta_x$ and $\eta_y$, the transverse electron distribution is changed to

$$g_1(p,X,Y) = \frac{N_0}{2\pi}\exp\left[-\frac{1}{2}(X - D_x p)^2\right]\exp\left[-\frac{1}{2}(Y - D_y p)^2\right], \tag{1}$$

where $D_x = \eta_x \sigma_\gamma / \sigma_x \gamma$ and $D_y = \eta_y \sigma_\gamma / \sigma_y \gamma$ is the dimensionless dispersive strength in horizontal and vertical, respectively. The transversely dispersed electron beam is then sent through a modulator to interact with a wave-front tilted seed laser pulse as shown in figure 1. Here we define a wave-front tilt gradient parameter of the seed laser as

$$\tau_s = \varphi/\sigma_s, \tag{2}$$

where $s$ represents for $x$ or $y$, $\varphi$ is the wave-front phase change over the transverse beam size of $\sigma_s$. After passing through the modulator, the electron beam is modulated with amplitude $A = \Delta\gamma/\sigma_\gamma$, where $\Delta\gamma$ is the energy modulation depth induced by the seed laser, and the dimensionless energy deviation of the electron beam becomes

$$p' = p + A\sin(k_s z + T_x X + T_y Y), \tag{3}$$

where $k_s$ is the wave number of the seed laser, $T_x = \tau_x \sigma_x$ and $T_y = \tau_y \sigma_y$ are the dimensionless wave-front tilt gradient parameters in horizontal and vertical. The three-dimensional distribution function of the electron beam after interacting with the seed laser can be written as

$$h_1(\zeta,p,X,Y) = \frac{N_0}{\sqrt{8\pi^3}} \exp\left\{-\frac{1}{2}\left[p - A\sin(\zeta - T_x X - T_y Y)\right]^2\right\} \exp\left[-\frac{1}{2}\left\{X - D_x\left[p - A\sin(\zeta - T_x X - T_y Y)\right]\right\}^2\right]$$
$$\times \exp\left[-\frac{1}{2}\left\{Y - D_y\left[p - A\sin(\zeta - T_x X - T_y Y)\right]\right\}^2\right] \quad (4)$$

where $\zeta = k_s z$ is the phase of the electron beam. After passing through the DS with the dispersive strength of $R_{56}$, the longitudinal beam distribution evolves to

$$h_2(\zeta,p,X,Y) = \frac{N_0}{\sqrt{8\pi^3}} \exp\left\{-\frac{1}{2}\left[p - A\sin(\zeta - T_x X - T_y Y - Bp)\right]^2\right\} \exp\left[-\frac{1}{2}\left\{X - D_x\left[p - A\sin(\zeta - T_x X - T_y Y - Bp)\right]\right\}^2\right]$$
$$\times \exp\left[-\frac{1}{2}\left\{Y - D_y\left[p - A\sin(\zeta - T_x X - T_y Y - Bp)\right]\right\}^2\right] \quad (5)$$

where $B = R_{56} k_s \sigma_\gamma / \gamma$ is the dimensionless strength of the DS. Integration of Eq (5) over $p$, $x$ and $y$ gives the beam density $N$ as a function of $\zeta$, $N(\zeta) = \int_{-\infty}^{\infty} dx \int_{-\infty}^{\infty} dy \int_{-\infty}^{\infty} dp\, h_2(\zeta,p,X,Y)$. And the bunching factor at $n$th harmonic can be written as

$$b_n = \frac{1}{N_0} \int_{-\infty}^{\infty} dp\, e^{-inp(T_x D_x + T_y D_y + B) - in(T_x X + T_y Y)} f_0(p) g_0(X,Y) \left\langle e^{-in(\zeta + AB\sin\zeta)} \right\rangle = J_n[nAB] e^{-(1/2)[n(T_x D_x + T_y D_y + B)]^2} e^{-(1/2)(nT_x + nT_y)^2}, \quad (6)$$

For the case without transverse dispersion and seed laser wave-front tilt, i.e. $D_x = D_y = 0, T_x = T_y = 0$, Eq. (6) reduces to the well-known formula for the bunching factor in a conversional HGHG FEL[11].

For the harmonic number $n > 4$, the maximal value of the Bessel function in Eq. (6) is about $0.67/n^{1/3}$ and is achieved when its argument is equal to $n + 0.81n^{1/3}$. For a given value of energy modulation amplitude $A$, the optimized strength of the DS is

$$B = (n + 0.81n^{1/3})/nA. \quad (7)$$

The maximal value of $b_n$ in Eq. (6) will be achieved when $T_x D_x + T_y D_y = -B$, which gives the optimized relation of $\tau$ and $\eta$:

$$\tau_x \eta_x + \tau_y \eta_y = -(n + 0.81n^{1/3})\gamma / nA\sigma_\gamma. \quad (8)$$

Notice that the third term in the right hand of Eq. (6) can be quite close to one when adopting a large $A$ and $\eta$ or a small horizontal beam size $\sigma_x$, so the maximal value of the nth harmonic bunching factor will approach

$$b_n \approx 0.67/n^{1/3}. \quad (9)$$

This is much larger than the bunching factor of a conversional HGHG, which exponentially decreases as the harmonic number increases.

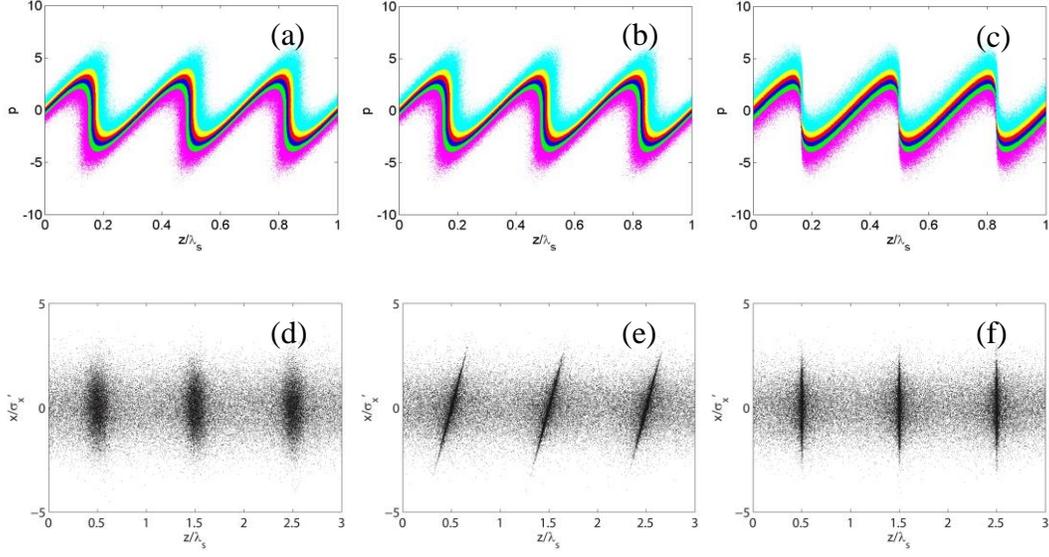

FIG.2. Comparison of beam longitudinal phase space (p-z) and normalized transverse electrons distributions (x-z) for different electron and laser beam injection conditions.

The physical mechanism behind the three-dimensional manipulation is the transverse-longitudinal phase space coupling. Comparisons of beam longitudinal phase spaces (p-z) and transverse electron beam distributions (x-z) for different electron and laser beam injection conditions are illustrated in Fig. 2. For simplicity, here we neglect the transverse emittance, assume $D_y = T_y = 0$ and only show the beam phase space within a longitudinal range of three seed laser wavelengths. The energy modulation amplitude is chosen to be $A = 3$ here, and the optimized condition for the horizontal wave-front tilt and dispersion strength is about $T_x D_x = -B \approx -0.35$ according to Eq. (7). Fig. 2 (a) shows the longitudinal phase space and transverse beam distribution after the DS for a conversional HGHG, the energy modulation is converted to density modulation and the localized current bumps (micro-bunching) that contain higher harmonic components are clearly seen in Fig. 2 (d). If the electron beam is transversely dispersed before the modulator and the seed laser is incident without any wave-front tilt, the final longitudinal phase space at the exit of the DS is shown in Fig. 2 (b), which is just like the phase space of the conversional HGHG. However, the transverse distribution for this case (Fig. 2 (e)) is quite different from that of HGHG (Fig. 2 (d)). One can observe that the electrons are concentrate to narrow slashes around the zero-phase of the seed laser, which implicates that the micro-bunching has already been significantly enhanced for high harmonics. However, the longitudinal dispersion results in different path lengths for electrons with different energies when passing through the DS. Thus the micro-bunching is tilt in the x-z space, which may result in a wave-front tilt of the output radiation. This micro-bunching tilt can be simply counteracted by turning the wave-front tilt of the seed laser, which changes the relative phase of the electrons with different transverse positions (different initial energies) in the modulator and so can be used to perform the phase-merging effect like in PEHG mechanisms[21-22]. After the DS, the micro-bunching will converge at a same phase as shown in Fig. 2 (c) and (f), which indicates a strong current modulation and a presence of strong components at higher harmonics of the seed wavelength.

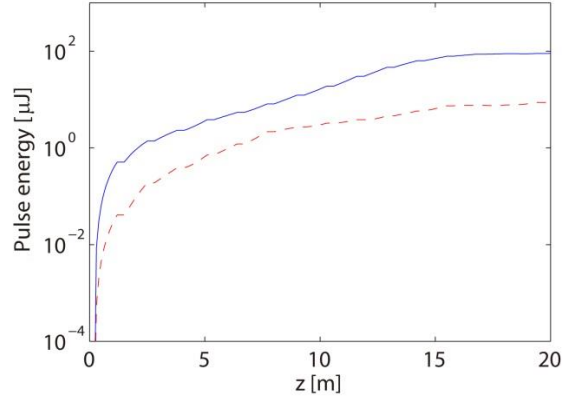

FIG.3. Evolution of the soft x ray FEL pulse energy along the radiator: comparison of the double-stage cascaded HGHG (red dashed line) and a single stage HGHG with three-dimensional phase space manipulation (blue solid line).

To illustrate a possible application of this three-dimensional manipulation technique with realistic parameters in a single stage seeded FEL, we take the nominal parameters of the Shanghai soft x ray FEL (SXFEL) test facility[23]. The SXFEL aims at generating 8.8 nm FEL from a 264 nm seed laser through a double-stage cascaded HGHG. The required electron beam energy is 840 MeV at the end of the linac. The bunch charge is about 0.5 nC. With these parameters, start-to-end tracking of the electron beam, including all components of SXFEL, has been carried out. The electron beam dynamics in photo-injector was simulated with ASTRA[24] to take into account space-charge effects. ELEGENT[25] was then used for the simulation in the remainder of the linac. The peak current of the electron beam is about 600 A at the exit of the linac. A constant profile is maintained in the approximately 600 fs wide and over 500 A region. A normalized emittance of approximately 0.65 mm-mrad and slice energy spread of about 100 keV are obtained. After horizontally dispersing the electron beam by the dogleg dispersion $\eta_x = 0.5m$, the average value of the horizontal beam size $\sigma_x$ is increased from about 60 μm to 70 μm, which will not significantly affect the FEL performance.

The three-dimensional phase space manipulation process and the FEL performance were simulated with GENESIS[26] based on the output of ELEGENT. For comparison purpose, simulations for double-stage cascaded HGHG have also been carried out. A 264 nm seed pulse with longitudinal pulse length of about 8 ps is adopted for single stage case and a much shorter pulse length of about 100 fs is adopted for the cascaded HGHG case. The length of the modulator is about 1 m with period length of 80 mm. To maximize the bunching factor at 30th harmonic of the seed laser, the parameters for the single stage HGHG case are set to be $A = 5$, $B = 0.2$ and $\tau_x \approx 3.6 rad/mm$. The bunching factor at the entrance of the radiator is about 5%. The period length of the radiator is 25 mm with K value of about 1.3. The evolution of the radiation pulse energy is shown in Fig. 3, where the large bunching factor at the entrance to the radiator offered by the three-dimensional phase space manipulation is responsible for the initial steep quadratic growth of the power. The significant output pulse energy enhancement is clearly seen in Fig. 3. The pulse energy of the 30$^{th}$ harmonic radiation approaches 100μJ at saturation, which is one order of magnitude higher than that of the original design of SXFEL with double-stage cascaded HGHG. Moreover, the 8.8nm radiation saturates within 15 m long undulator, which is in the range of original design of SXFEL.

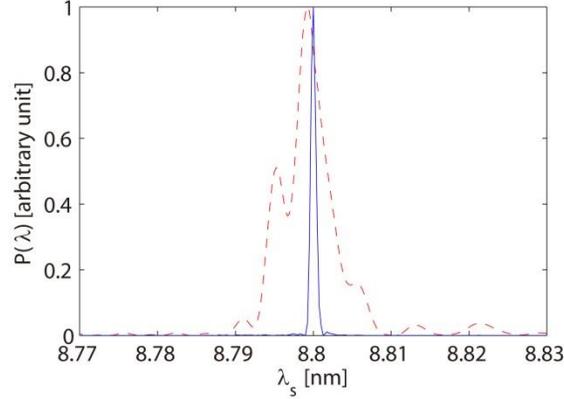

FIG.4. Comparison of spectra at saturation for double-stage cascaded HGHG (red dashed line) and a single stage HGHG with three-dimensional phase space manipulation (blue solid line).

The single-shot radiation spectra for these two cases at saturation are shown in Fig. 4. The spectral bandwidth at saturation for the single stage HGHG with three-dimensional phase space manipulation (blue solid line) is about 0.08%, which is quite close to transform limit. However, the bandwidth of the cascaded HGHG (red dashed line) is about 0.6%, which is 1.7 times wider than the transform limit. The bandwidths broadening and spectral noises in the spectrum of cascaded HGHG are mainly induced by the initial energy curvature of the electron beam[27, 28], which is unavoidable in the FEL linacs due to the radio frequency curvature and wake field effects. However, the dependence of output spectrum in seeded FEL on the residual beam energy chirp can be canceled when using the three-dimensional phase space manipulation technique. Here we assume an initial linear energy chirp $h$ in the electron beam. After transversely dispersing the electron beam with $D_x$, the longitudinal energy chirp results in a similar scaled chirp in the electron beam transverse distribution:

$$X' = X + D_x(p + h\zeta). \tag{10}$$

It has been point out that the electrons with different transverse positions will get different energy modulation phases in the modulator when the seed laser wave-front is tilted by $T_x$. Here we plug Eq. (10) into Eqs. (1-6) and get the bunching factor for an electron beam with linear chirp as

$$b_n = \frac{1}{N_0} \int_{-\infty}^{\infty} dp\, e^{-inp(T_x D_x + B) - in T_x X} f_0(p) g_0(X,Y) \left\langle e^{-in\zeta[1+h(T_x D_x + B)]} e^{-inAB\sin\zeta} \right\rangle. \quad (11)$$

So the optimized output frequency can be written as

$$k_n = \frac{n k_s}{1 + h(T_x D_x + B)}, \quad (12)$$

As the optimized condition for the bunching factor is $T_x D_x = -B$, one can easily find in Eq. (12) that the output radiation wavelength and bandwidth is naturally immune to the beam energy chirp for the single-stage HGHG with three-dimensional phase space manipulation.

In summary, we have proposed a simple method for three-dimensional manipulation of the electron beam phase space in the optical scale. Analytical and numerical investigations of the proposed method demonstrate a great potential of generating high harmonic radiation with a relatively small energy modulation and a suitable transverse dispersion in a single stage seeded FEL. Just like the PEHG, the transverse dispersion cannot be closed after the density modulation, which will cause FEL gain reduction compared with the conversional HGHG due to the increased beam size in the radiator[21, 22]. However, considering that the ability of exploiting the full electron bunch and naturally immune to the initial energy curvature in the electron beam, the output pulse energy and the bandwidth will be significantly improved by using this method, and thus leads a FEL average bandwidth brightness nearly 2 orders of magnitude higher than the conversional double-stage HGHG baseline for a seeded soft x-ray FEL facility.

In addition to generation of fully coherent radiation at soft x-ray, intensity studies for other possible applications of three-dimensional phase space manipulation have also been carried out, and the results indicate that this method can be used for ultra-high harmonic and ultra-short pulses generation. The concept of wave-front tilt seed laser can also be used to significantly improve the performance of direct seeding FELs based on the laser-plasma accelerators.

The authors would like to thank B. Liu and D. Wang for helpful discussions and useful comments. This work is supported by the Major State Basic Research Development Program of China (2011CB808300) and the National Natural Science Foundation of China (11175240, 11205234 and 11322550).